\documentclass[%
 reprint,
showpacs,
 amsmath,amssymb,
 aps,
 pra,
]{revtex4-1}

\usepackage[cmyk]{xcolor}

\usepackage{graphicx} 
\usepackage{dcolumn} 
\usepackage{bm} 

\usepackage[utf8]{inputenc}
\usepackage[T1]{fontenc}
\usepackage{times}
\usepackage{flafter} 
\usepackage{textgreek} 
\usepackage{mathrsfs} 

\newcommand{\ii}{\mathrm{i}}
\newcommand{\ee}{\mathrm{e}}

\renewcommand{\Re}{\operatorname{Re}}

\newcommand{\fullwf}{\mathit{\Phi}}
\newcommand{\velel}{{v_\mathrm{ee}}}

\newcommand{\pr}{^\prime}
\newcommand{\conj}{^\ast}

\newcommand*\diff{\mathop{}\!\mathrm{d}}
\newcommand{\derivative}{\partial}

\newcommand{\abs}[1]{\left|#1\right|}
\newcommand{\iinteg}[3]{\iint_{#3}\!\!\mathrm{d}#1\mathrm{d}#2\,}
\newcommand{\integ}[3]{\int_{#2}^{#3}\!\!\mathrm{d}#1}
\newcommand{\iek}[1]{\left(#1\right)}

\newcommand{\rno}{\tilde{\phi}}
\newcommand{\rnou}{\underline{\tilde{\phi}}}
\newcommand{\ug}{\underline{\tilde{\gamma}}}
\newcommand{\ketrno}[1]{|\tilde{#1} \rangle}

\newcommand{\mat}[1]{\mathbf{\mathcal{#1}}}

\makeindex

\begin{document}

\title{Nonsequential double ionization with time-dependent renormalized-natural-orbital theory}

\author{M.\ Brics}
\author{J.\ Rapp}
\author{D.\ Bauer}%
\affiliation{%
 Institut für Physik, Universität Rostock, 18051 Rostock, Germany
}

\date{\today}

\begin{abstract}
Recently introduced time-dependent renormalized-natural-orbital theory (TDRNOT) is tested on nonsequential double ionization (NSDI) of a numerically 
exactly solvable one-dimensional model He atom subject to few-cycle, 800-nm laser pulses.
NSDI of atoms in strong laser fields is a prime example of non-perturbative, highly correlated electron dynamics. As such, NSDI is an important ``worst-case'' benchmark for any time-dependent few and many-body technique beyond linear response. It is found that TDRNOT reproduces the celebrated NSDI ``knee,'' i.e., a many-order-of-magnitude enhancement of the double ionization yield (as compared to purely sequential ionization) with only the ten most significant natural orbitals (NOs) per spin. Correlated photoelectron spectra---as ``more differential'' observables---require more NOs.

\end{abstract}


\maketitle

\section{Introduction}
Nonsequential double ionization (NSDI) in intense laser pulses has been experimentally observed in measurements of ion yields as a function of the 
laser intensity, which  deviate from the yields expected from a sequential ionization scenario, forming the so-called NSDI ``knee'' (see 
\cite{WBeckerRev,Carla2011} for recent reviews). In fact, the multiple ionization yields are typically enhanced by several  orders of magnitude.
With ionization yields being rather integrated observables the mechanism behind NSDI could not be unequivocally resolved until the measurement of ion spectra \cite{weberNSDI2000,moshammerNSDI2000} and correlated photoelectron distributions has become feasible
(see \cite{webernature2000,feuerstein2001} for early and, e.g., \cite{kuebel} for very recent work). 
Meanwhile NSDI is understood in terms of a recollision process: one electron is emitted but oscillates back to its parent ion to knock out the next 
electron. If the return energy is not sufficient for collisional ionization, the next electron might be excited and later emitted owing to the laser 
field \cite{RESI}. 

The described recollision scenario poses a huge challenge for general many-body methods when applied to such a few-electron test case. For example, in time-dependent Hartree-Fock (TDHF) or time-dependent density functional theory (TDDFT) applied to He starting from the singlet ground state there is only one spatial orbital describing both electrons (one spin-up, the other spin-down). Not surprisingly, it was found that such methods are not capable of describing NSDI \cite{Dahlen_Leeuwen}, although formally for different reasons. TDHF, as a mean-field approach, does not incorporate correlation by definition. TDDFT is in principle exact but only in the sense that it gives the exact time-dependent  electron density. However, even if the exact time-dependent  electron density was known from a TDDFT calculation employing the exact exchange-correlation potential \cite{MLein2005}, the exact double-ionization probability could still not be calculated because this observable is unknown as an explicit functional of 
the electron density, and simple approximations to it do not reproduce the NSDI knee \cite{petersilka,florian}. 

Solving the full time-dependent Schr\"odinger equation (TDSE) for He in full dimensionality and in strong, long-wavelength (i.e., $\geq 800$\,nm) 
laser fields is still beyond what is possible with current super computer technology \cite{KenTaylor2006}.  Therefore it is essential to develop 
practicable time-dependent many-electron methods beyond linear response that account for correlation. Time-dependent configuration interaction (TDCI) 
\cite{Rohringer,Santra},  multiconfigurational time-dependent Hartree (MCTDH) \cite{mctdh} or multiconfigurational TDHF (MCTDHF) 
\cite{Zanghellini,caillat,Scrinzi2006,mctdhf} are exact in principle. However,  the crucial question in practice is how many configurations or 
determinants are required to recover a certain strong correlation feature  such as the NSDI knee. General conclusions are difficult to draw, as 
different TDCI and MCTDHF approaches may vary in the single-particle basis functions chosen. It has been shown that for He (or H$_\mathrm{2}$) a 
time-dependent variational 
approach using a wavefunction ansatz with two different single-particle orbitals
 (time-dependent extended Hartree-Fock) \cite{Dahlen_Leeuwen,Bandrauk2006} or an {\em ad hoc} ansatz with an ``inner'' and an ``outer'' outer 
orbital \cite{Watson1997} at least generate kneelike structures in the double-ionization yield. However, they are only in poor agreement with   the 
exact numerical results available for low-dimensional models. To the best of our knowledge there are no systematic tests of computational approaches 
that demonstrate a convergence toward the exact NSDI knee. In fact, we are not aware of any work that accurately reproduces the NSDI knee using a 
many-body method that overcomes the ``exponential wall'' \cite{KohnNobel1999}.    In this work, we will provide such an analysis for our recently 
introduced time-dependent renormalized-natural-orbital theory (TDRNOT) \cite{tdrnot,tdrnot2}.


\section{Theory}
Before we present results on the NSDI knee (Sec.~\ref{sec:ioniyields}) and on correlated photoelectron spectra (Sec.~\ref{sec:temd}), we briefly  
introduce the He model, review the essentials of TDRNOT, particularly when applied to a two-electron system, and discuss the observables to be 
calculated.

Atomic units (a.u.) are used unless otherwise indicated. 

\subsection{Model atom}
The widely applied one-dimensional helium model atom \cite{Dieter97,Lappas_Leeuwen,Lein2000,Dahlen_Leeuwen,MLein2005,florian,florian2,mctdhf} in a laser field  has the Hamiltonian 
\begin{align}
  \hat H^{(1, 2)}(t)
    &=
      \hat h^{(1)}(t)
      + \hat h^{(2)}(t)
      + \velel^{(1, 2)}, \label{hamiltonian}
\end{align}
where upper indices indicate the action on either electron $\mathrm 1$, electron $2$, or both. The single-particle Hamiltonian reads $\hat 
h(t)=\hat{H}_A+\hat{H}_L(t)$,  with
\begin{align}
  \hat{H}_A &=
      \frac{\hat p^2}{2}
      - \frac{2}{\sqrt{x^2 + \varepsilon_{\mathrm{ne}}}},
\end{align}
$\hat{H}_L(t)=A(t)\hat p$ (dipole approximation and velocity gauge with the $A^2$ term transformed away),
 and the electron-electron interaction 
\begin{align}
  \velel^{(1, 2)}
    &=
      \frac{1}{\sqrt{\left(x^{(1)}-x^{(2)}\right)^2 + \varepsilon_\mathrm{ee}}}\,.
\end{align}
 The electron-ion smoothing parameter 
$\varepsilon_\mathrm{ne}=0.50$ 
is chosen such that the groundstate energy of  $\mathrm{He}^+$  $E_0^\mathrm{He^+}=-2.0$ is recovered. The electron-electron smoothing parameter 
$\varepsilon_\mathrm{ee}=0.33$ is 
tuned to yield  the neutral-He energy $E_0^\mathrm{He}=-2.9$.

\subsection{Density matrices, renormalized-natural-orbitals, and their equations of motion}
The Hamiltonian (\ref{hamiltonian}) does not act on the spin, which---in the two-particle case---allows one to factorize the wavefunction,
\begin{equation}
  \langle 12|\fullwf (t)\rangle
    =
  \fullwf(12;t)
    =
      \fullwf(x_1 x_2;t)
      \fullwf_{\sigma_1 \sigma_2}. \label{wf-factorization}
\end{equation}
Here $1$ and $2$ are shorthand notations for position and spin $(x_1, \sigma_1)$ and $(x_2, \sigma_2)$, respectively.
The two- and one-body density matrices read
\begin{align}
  \gamma_2(12, 1\pr2\pr;t)
    &=
      \fullwf(12;t)
      \fullwf\conj(1\pr2\pr;t),
\end{align}
\begin{align}
  \gamma_1(1, 1\pr;t)
    &=
      2\int\diff 2\,
      \gamma_2(12, 1\pr2;t).
\end{align}
Natural orbitals (NOs) $ \phi_k(1;t)=\langle 1 | k(t) \rangle$ are defined as eigenvectors of $\gamma_1$:
\begin{align}
  \gamma_1(1, 1\pr;t)
    &=
      \sum_k
      n_k(t)
      \phi_k(1;t)
      \phi_k\conj(1\pr;t).
\end{align}
The corresponding eigenvalues $n_k(t)\in\left[0, 1\right]$ are called occupation numbers (ONs). NOs and ONs were introduced a long time ago (see, 
e.g., \cite{loewdin,coleman,colemanorangebook}), but only recently have their usefulness for time-dependent few- and many-body problems
been studied \cite{Pernal,appel,giesbertz,appelgross,helbig,giesbjcp2012}.

The coupled equations of motion for the ONs and the NOs can be unified by introducing renormalized NOs (RNOs) \cite{tdrnot} 
\begin{equation}
\langle 1| \tilde k(t)\rangle = \rno_k(1;t)=\sqrt{n_k(t)}\phi_k(1;t)
\end{equation} 
so that
\begin{equation}
n_k(t)=\langle \tilde k(t)| \tilde k(t)\rangle 
\end{equation} 
and
\begin{align}
  \gamma_1(1, 1\pr;t)
    &=
      \sum_k
      \tilde \phi_k(1;t)
      \tilde \phi_k\conj(1\pr;t) .
\end{align}
The two-body density matrix expanded in RNOs reads
\begin{equation}
\begin{split}
  &\gamma_2(12, 1\pr 2\pr;t)\\
    =&
      \sum_{ijkl}
      \tilde \gamma_{2,ijkl}(t)
      \rno_i(1;t)
      \rno_j(2;t)
      \rno_k\conj(1\pr;t)
      \rno_l\conj(2\pr;t) .
      \end{split}
\end{equation}

The equation of motion (EOM) for the RNOs  is \cite{tdrnot2}
\begin{equation}
\begin{split} 
  \ii\derivative_t\ketrno{n}
    &=
      \hat h(t)\ketrno{n}+ {\mat{A}}_n(t) \ketrno{n}\\
       & \qquad + \sum_{k\neq n} {\mat{B}}_{nk}(t)\ketrno{k}
       + \sum_k {\mat{\hat{C}}}_{nk} (t) \ketrno{k} \label{eom-rno}
\end{split}
\end{equation}
with 
\begin{equation}
  \mat{{A}}_n(t)
    =
      -\frac{1}{{n}_n(t)}\Re\sum_{jkl}
      \tilde\gamma_{2,njkl}(t)
      \langle \tilde k \tilde l |\velel|\tilde n \tilde j \rangle, \label{eom-rno-A}
\end{equation}
\begin{align}
\begin{split}
 {\mat{B}}_{nk}(t)
    &=
      \frac{2}{{n}_k(t)-{n}_n(t)}\sum_{jpl}\Bigl[
        \tilde \gamma_{2,kjpl}(t)
        \langle \tilde p \tilde l |\velel| \tilde n\tilde j \rangle \Bigl.\\
        & -\Bigr. \tilde\gamma_{2,plnj}(t)
        \langle \tilde k \tilde j |\velel| \tilde p \tilde l \rangle
      \Bigr],  \label{eom-rno-B}
\end{split}
\end{align}
and
\begin{align}
 {\mat{\hat{C}}}_{nk}(t)
    &=
      2\sum_{jl} \tilde \gamma_{2,kjnl}(t)\langle \tilde l|\velel| \tilde j \rangle. \label{eom-end}  
\end{align}
One observes that the effective Hamiltonian in the TDSE-like equation \eqref{eom-rno} consists of the usual one-body operator $\hat h(t)$, a diagonal 
part $ \mat{{A}}_n(t)\in\mathbb{R}$, the  part ${\mat{B}}_{nk}(t) \in \mathbb{C}$ which couples RNOs, and the operator ${\mat{\hat{C}}}_{nk}(t)$, 
which also couples RNOs. As the effective Hamiltonian in \eqref{eom-rno} is Hermitian, the corresponding time evolution of the RNOs is unitary. 

In general, there are infinitely many NOs required to describe a correlated quantum system, even if it contains only two particles. Ordered decreasingly according to their ONs, the number of RNOs taken into account in an actual numerical implementation of \eqref{eom-rno} is necessarily truncated, which introduces errors in the propagation. The effect of this truncation will be seen in the results in Sec.~\ref{sec:results} below.

\medskip

In the two-particle case  the expansion coefficients $\tilde \gamma_{2,ijkl}(t)$ are exactly known \cite{tdrnot2},
\begin{equation}
\tilde \gamma_{2,ijkl}(t)=(-1)^{i+k}\frac{e^{\ii[\varphi_i-\varphi_k]}}{2\sqrt{n_i(t)n_k(t)}}\delta_{i, j\pr}\delta_{k, l\pr}. \label{eq:gamma2tilde}
\end{equation}
Here, the ``prime operator'' acts on the positive integer $k$ according to
\begin{equation}
 k\pr = \begin{cases}
         k+1 &\mbox{if $k$ odd}\\
         k-1 & \mbox{if $k$ even,}
        \end{cases}
\end{equation}
and the phase factors are \cite{tdrnot2}
\begin{equation}
 e^{\ii\varphi_i^{(\mathrm S)}} = 
         2\delta_{k,1}+2\delta_{k,2}-1, \qquad e^{\ii\varphi_i^{(\mathrm T)}} = 1
\end{equation}
in the spin-singlet and -triplet case, respectively.
Note that 
the EOM for the RNOs \eqref{eom-rno} is given here for phase-including NOs \cite{giesbertz} so that $\varphi_i$ and $\varphi_k$ in  \eqref{eq:gamma2tilde} do not depend on time.
Employing the factorization (\ref{wf-factorization}) we can write
\begin{align}
  \gamma_2(12, 1\pr2\pr;t)
    &=
             \fullwf(x_1 x_2;t)
        \fullwf\conj(x_1\pr x_2\pr;t)
      \fullwf_{\sigma_1 \sigma_2}
      \fullwf\conj_{\sigma_1\pr \sigma_2\pr},
\end{align}
\begin{align}
  \gamma_1(1, 1\pr;t)
    &= \gamma_1(x_1, x_1\pr;t)
      \sum_{\sigma_2}
      \fullwf_{\sigma_1 \sigma_2}
      \fullwf\conj_{\sigma_1\pr \sigma_2}
\end{align}
where
\begin{align}
\gamma_1(x_1, x_1\pr;t)  &=  2\int\diff x_2\,
        \gamma_2(x_1 x_2, x_1\pr x_2;t)  \nonumber \\
    &=
      \sum_k \underline{n}_k(t) 
      \underline{{\phi}}_k(x_1;t)
      \underline{\phi}_k\conj(x_1\pr;t) \nonumber \\
&=
\sum_k 
      \rnou_k(x_1;t)
      \rnou_k\conj(x_1\pr;t),\\
\gamma_2(x_1 x_2, x_1\pr x_2\pr;t)  &= 
        \fullwf(x_1 x_2;t)
        \fullwf\conj(x_1\pr x_2\pr;t) \nonumber \\
    &=
      \sum_{ijkl}
      \ug_{2,ijkl}(t)
      \rnou_i(x_1;t)
      \rnou_j(x_2;t) \nonumber \\
&\qquad\quad
    \times  \rnou_k\conj(x_1\pr;t)
      \rnou_l\conj(x_2\pr;t).
\end{align}
Here and in the following, spatial RNOs and quantities calculated from them (e.g., $\underline{n}_i(t)=\langle \rnou_i(t) |\rnou_i (t) \rangle$) will  be indicated by underlining them.
How the RNOs can be written as a factorization in the spatial and the spin part is discussed in detail in 
 \cite{tdrnot2}. In this work we will only consider results for the singlet configuration where  the RNOs with $k=1,2,3, \ldots$ can be arranged as
\begin{equation}
\langle x | \tilde  k(t)\rangle =  \left\{ \begin{array}{ll} |\!\uparrow\rangle\,\rnou_{k'/2}(x;t) & \mathrm{if}\ k\ \mathrm{odd} \\   
|\!\downarrow\rangle\, \rnou_{k/2}(x;t) & \mathrm{if}\  k\ \mathrm{even} \end{array} \right. 
\end{equation}
so that any consecutive $k$-odd and $k+1$-even RNOs share the same spatial component $\rnou_{k'/2}(x;t)$.

\subsection{Observables} 
\label{sec:observables}
We are interested in  the double-ionization probability  of the model He atom as a function of the laser intensity and in correlated photoelectron spectra, i.e., the probability to find one electron being emitted with momentum $p_1$ and the other with $p_2$, for laser intensities where NSDI occurs. Both should in principle be calculated via the  projection of the wavefunction after the laser pulse on two-electron continuum states of asymptotic momenta $p_1$ and $p_2$. However, this approach is numerically unfeasible. We will shortly explain how the yields and spectra are calculated in a less rigorous but sufficiently accurate manner in this work. 

\subsubsection{Ionization probabilities}
An efficient way to calculate ionization probabilities from the two-electron wavefunction $\fullwf(x_1 x_2)$ after the laser pulse  is based on the integration  of the probability density $\abs{\fullwf(x_1 x_2)}^2$ over certain  spatial regions,
\begin{align}
P^{0}
    =
      &\iinteg{x_1}{x_2}{\abs{x_1},\abs{x_2}<a}\,\abs{\fullwf(x_1 x_2)}^2,  \label{eq:P0_psi}\\
  P^{2+}
    =
      &\iinteg{x_1}{x_2}  {\abs{x_1},\abs{x_2}\geq a}\,\abs{\fullwf(x_1 x_2)}^2 ,  \label{eq:P2_psi}\\
P^{1+} = &1 - P^{0} -  P^{2+}
\end{align}
where we made use of the fact that $P^{0}+P^{1+}+P^{2+}=1$.
The parameter $a>0$ should be sufficiently large such that the probabilities $P^{1+}$ and $P^{2+}$ are negligible for the groundstate
and singly-excited eigenstates.
On the other hand,  $a$ should not be too large so that the probability density describing ionization does not need too much time to leave the neutral-He region $\abs{x_1},\abs{x_2}<a$.
For our model we chose $a=6$.

As for a two-electron system $\abs{\fullwf(x_1 x_2)}^2=\gamma_2(x_1 x_2, x_1 x_2)$,  Eqs.~\eqref{eq:P0_psi} and \eqref{eq:P2_psi} are read in terms  
of RNOs:
\begin{align}
\begin{split}
P^{0}=
    &\sum_{ijkl}
      \ug_{2,ijkl}      
      \integ{x_1}{-a}{a}\,\rnou_i(x_1)\rnou_k\conj(x_1) \\ 
  &\times \integ{x_2}{-a}{a} \,
      \rnou_j(x_2) \rnou_l\conj(x_2) , \label{eq:P0_no}
\end{split}\\
\begin{split}  
P^{2+}=
    &\sum_{ijkl}
      \ug_{2,ijkl}
      \integ{x_1}{\abs{x_1}\geq a}{}\, \rnou_i(x_1)\rnou_k\conj(x_1) \\
 &\times \integ{x_2}{\abs{x_2}\geq a}{}\,\rnou_j(x_2) \rnou_l\conj(x_2).  \label{eq:P2_no}
\end{split}
\end{align} 
Note that in the two-electron case this form is equivalent to first reconstructing the wavefunction (which is possible for two electrons 
\cite{tdrnot2}) and then using  Eqs.~\eqref{eq:P0_psi} and \eqref{eq:P2_psi}.

\subsubsection{Momentum distributions}
\label{sec:twe_em_mom_ditr}

A numerically efficient method to calculate correlated double-ionization photoelectron spectra  is to multiply  the two-electron wavefunction by a 
mask function $f(x_1 x_2)$, which removes the parts representing He$^+$ and neutral He: $$\fullwf^{2+}(x_1 x_2) \simeq f(x_1 x_2)\fullwf(x_1 x_2).$$
Here, we chose $f(x_1 x_2)=f(x_1)f(x_2)$, with $f(x)=1/\sqrt{1+\ee^{-c(\abs{x}-a)}}$ and $c=1.25$ \cite{florian2}. After Fourier transforming 
$\fullwf^{2+}(x_1 x_2)$  to momentum space,
\begin{equation}
 \fullwf^{2+}(p_1 p_2)=\frac{1}{2\pi}\!\integ{x_1}{}{}\integ{x_2}{}{}\, \fullwf^{2+}(x_1 x_2)\, e^{-\ii(p_1 x_1+p_2 x_2)} ,
\end{equation}
the  double-ionization photoelectron spectrum is obtained as  $$\rho^{2+}(p_1 p_2)=2\abs{\fullwf^{2+}(p_1 p_2)}^2.$$

In our TDRNOT treatment we proceed analogously by first defining 
\begin{equation}
\gamma^{2+}_2(x_1 x_2,x_1\pr x_2\pr) = f(x_1 x_2)f\conj(x_1\pr x_2\pr) \gamma_2(x_1 x_2,x_1\pr x_2\pr),
\end{equation}
whose Fourier transform  is $\gamma^{2+}_2(p_1 p_2,p_1\pr p_2\pr)$. Then, $$\rho^{2+}(p_1 p_2)=2 \gamma^{2+}_2(p_1 p_2,p_1 p_2),$$  which can be 
written as
\begin{equation}
\begin{split}
 &\rho^{2+}(p_1 p_2) \\ &\simeq 2 \sum_{ijkl}
      \ug_{2,ijkl}  \rnou_i^{+}(p_1)\rnou_j^{+}(p_2)\left\{\rnou_k^{+}(p_1)\rnou_l^{+}(p_2)\right\}\conj 
\end{split}
\end{equation}
where
\begin{align}
  \rnou_i^{+}(p_j)&=\frac{1}{\sqrt{2 \pi}} \integ{x_j}{}{}\, f(x) \rnou_i(x_j) \, e^{-\ii x_jp_j}.
\end{align}
We thus have an explicit construction for $\rho^{2+}(p_1 p_2)$ in terms of RNOs. Note that in TDDFT such a construction in terms of Kohn-Sham orbitals is unknown \cite{florian2}.

\begin{figure}[htbp]
\includegraphics[width=0.9\columnwidth]{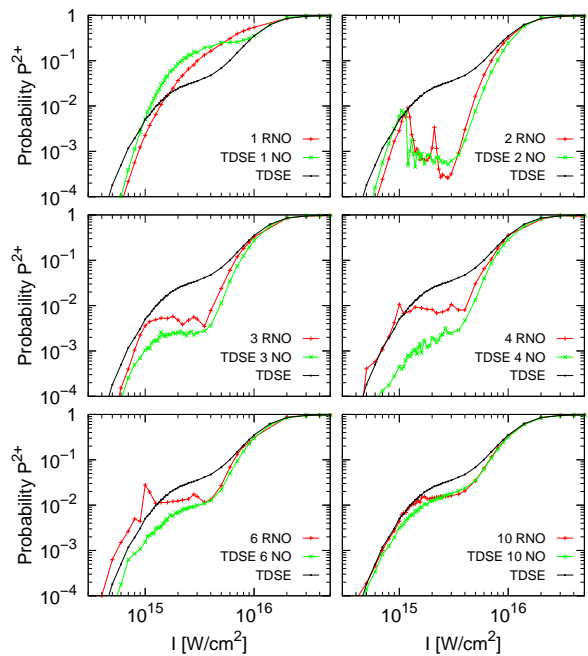}
\caption{(Color online) Double-ionization probability vs\ laser intensity. TDRNOT results with $N=1,2, 3, 4, 6, 10$  spatial RNOs (red, +) are 
compared with the exact  TDSE result (black, dots)  and with  the ionization 
probability reconstructed using the first $N$ exact NOs calculated from the exact TDSE wavefunction (green, $\times$).}
\label{fig:p2}
\end{figure}

\section{Results and discussion} \label{sec:results}
We consider an $800$-nm ($\omega=0.058$) linearly 
polarized $N_\mathrm{cyc}=3$-cycle  $\sin^2$-shaped laser pulse  of duration $T=2\pi N_\mathrm{cyc}/\omega$. The vector potential in dipole approximation reads
\begin{equation}
  A(t)=\hat{A}\sin^2\iek{\frac{\omega t}{2N_\mathrm{cyc}}}\sin(\omega t) \qquad \text{for }\  0\leq t \leq T 
\end{equation}
and zero otherwise. The numerical grids for both the TDSE-benchmark and TDRNOT calculations covered $\pm 1500$~a.u.\ in the spatial directions.

\subsection{Ionization yields} \label{sec:ioniyields}
Figure~\ref{fig:p2} shows the double-ionization probability $P^{2+}$ as a function of the laser intensity for different numbers of spatial RNOs $N$ 
between $1$ (upper-left panel) and $10$ (lower-right panel). For comparison, the exact TDSE result is included in black in all panels. The 
nonmonotonic behavior of the first derivative of this exact $P^{2+}$ curve in the region around $2\times 10^{15}$\,W/cm$^2$ gives rise to the 
celebrated NSDI knee. 

A TDRNOT calculation with $N=1$ RNO per spin yields a featureless $P^{2+}$ curve, as seen in the upper-left panel of Fig.~\ref{fig:p2}. In fact, in 
the case of a two-electron spin-singlet system, a single NO per spin is equivalent to TDHF or TDDFT in exchange-only approximation, for which it is 
already known that the NSDI knee is not reproduced \cite{Dahlen_Leeuwen,MLein2005,petersilka,florian}.

Truncating the number of RNOs in a TDRNOT calculation introduces an error in the propagation of the RNOs \cite{tdrnot2}. This error should be distinguished from the error that arises alone due to the fact that only a finite number of NOs is taken into account for the calculation of an observable. We do this by determining all exact NOs from the exact TDSE wavefunction but consider only the $N$ most dominant of them to calculate the observable $P^{2+}$. The respective results are also shown in  Fig.~\ref{fig:p2}. For $N=1$ this procedure gives a result very different from the  TDRNOT with $N=1$. There is even already a knee in the TDSE-1-NO result, albeit a quantitatively wrong one.   Both TDRNOT with a single RNO and the TDSE-1-NO curve show a wrong slope in the limit of low laser intensity.

For $N=2$ NOs per spin  (upper-right panel) a knee appears also in the TDRNOT result. It is exaggerated and jaggedly structured, and underestimates 
the  $P^{2+}$ yield. A similar behavior with two orbitals was observed in extended Hartree-Fock treatments \cite{Dahlen_Leeuwen,Bandrauk2006} and 
with the so-called  ``crapola'' model \cite{Watson1997}, where an ``inner'' and an ``outer'' orbital is postulated. 

With increasing $N$ the agreement between TDRNOT results and TDSE improves. For $N=10$ the truncation error in the propagation of the RNOs is small 
enough to give almost the same probability 
 $P^{2+}$ as if it was calculated with the first $N=10$ exact NOs. 

\begin{figure}[htbp]
\includegraphics[width=0.8\columnwidth]{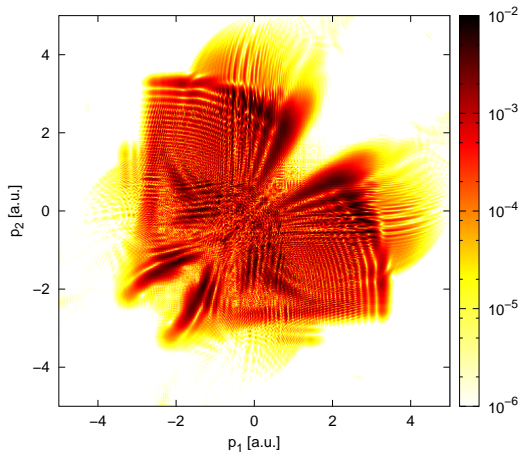}
\caption{(Color online) TDSE-benchmark two-electron photoelectron spectrum $\rho^{2+}(p_1 p_2)$ at $I=2.25\times 10^{15}\; \mathrm{W/cm^2}$. 
\label{fig:p2dim_tdse}}
\end{figure}

Clearly, our TDRNOT approach is only attractive if $N$ can be kept reasonably small. We have shown in Refs.~\cite{tdrnot,tdrnot2} how, with a few 
RNOs, doubly excited states, autoionization, and  Rabi flopping can be described using TDRNOT. Unfortunately, NSDI is more demanding in $N$, meaning 
that NSDI is highly correlated, and thus many more NOs than particles are required. Moreover, note that although NSDI is a huge effect on the 
$P^{2+}$ level, it is a small effect compared to the probability for single-ionization $P^{1+}$, and small effects on an absolute scale are captured 
by NOs with small ONs. The dominant NOs are mainly ``responsible'' for single ionization, or no ionization at all. In that respect it is remarkable 
to achieve an agreement such as the one shown for $N=10$ spatial RNOs in  Fig.~\ref{fig:p2}.  We are not aware of any TDCI or TDMCHF calculation that 
achieved such an agreement, let alone with only ten basis functions.

\begin{figure}[htbp]
\includegraphics[width=0.8\columnwidth]{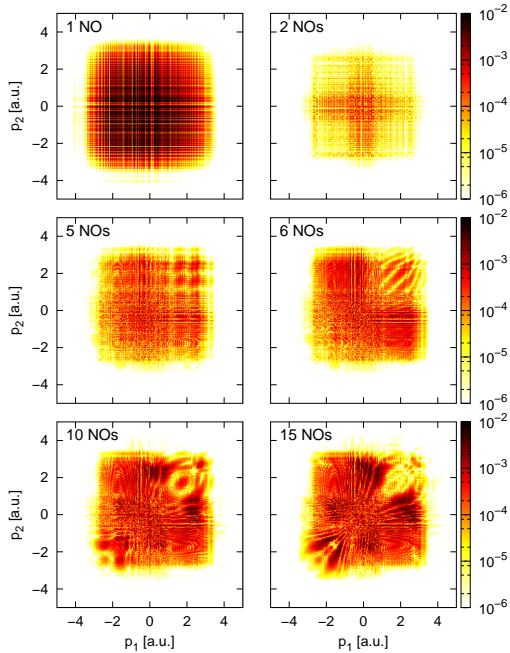}
\caption{(Color online) $\rho^{2+}(p_1 p_2)$   at $I=2.25\times 10^{15}\; \mathrm{W/cm^2}$ obtained from the $N=1,2,5,6,10,15$ dominant, exact spatial NOs calculated from the exact TDSE wavefunction after the laser pulse.}
\label{fig:p2dim}
\end{figure}

\subsection{Two-electron momentum distribution} \label{sec:temd}
Correlated photoelectron spectra contain more information than ionization probabilities. In general, the ``more differential'' an observable is, the harder it is to reproduce by some approximate method because the dynamic range to be  accurately covered increases. An additional, conceptual challenge arises with TDDFT because $\rho^{2+}(p_1 p_2)$ is an unknown functional of the single-particle density, and simple approximations fail \cite{florian2}.

Figure~\ref{fig:p2dim_tdse} shows the TDSE benchmark result for $\rho^{2+}(p_1 p_2)$ at $I=2.25\times 10^{15}\; \mathrm{W/cm^2}$, i.e., in the NSDI intensity regime. The butterfly structure indicating electrons emitted into the same direction is characteristic of NSDI \cite{WBeckerRev,Carla2011} and has been essential to identify rescattering as its origin.

From the TDSE benchmark we know that the first thousand exact NOs have ONs $> 10^{-15}$. The question is how many NOs are needed to recover the butterfly structure seen in  Fig.~\ref{fig:p2dim_tdse}.
Figure~\ref{fig:p2dim} shows that with the first $15$ exact NOs from the TDSE simulation the butterfly structure  of  Fig.~\ref{fig:p2dim_tdse} 
emerges, but details are still not accurately represented over the 4 orders of magnitude dynamic range shown. However, it is sufficient for the 
purpose of validating TDRNOT with a reasonably small number of RNOs. Up to $N=5$, mainly uncorrelated, gridlike horizontal and vertical structures 
are visible. From $N=6$ on, however, clear correlated structures appear, first in the first quadrant $p_1, p_2 > 0$.

\begin{figure}[htbp]
\includegraphics[width=0.8\columnwidth]{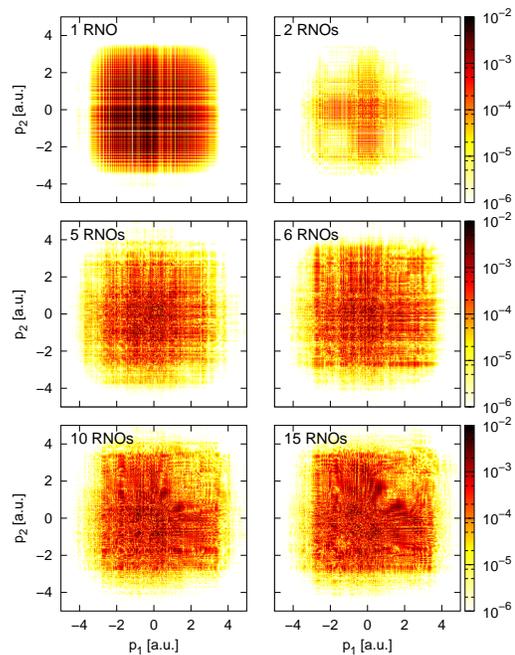}
\caption{(Color online) $\rho^{2+}(p_1 p_2)$   at $I=2.25\times 10^{15}\; \mathrm{W/cm^2}$ obtained from TDRNOT with $N=1,2,5,6,10,15$ spatial RNOs.}
\label{fig:p2dim_TDROT}
\end{figure}

Figure~\ref{fig:p2dim_TDROT} shows the corresponding TDRNOT result with $N$ RNOs per spin propagated.  Again, the differences between the benchmark 
results in Fig.~\ref{fig:p2dim}  and TDRNOT in Fig.~\ref{fig:p2dim_TDROT} are due to the truncation error in the number of propagated RNOs. This 
truncation error severely spoils the correlation structure in the first quadrant; only for $N=15$ does it start to emerge. In order to reproduce, 
say, the lower-right spectrum in Fig.~\ref{fig:p2dim}, one would need to propagate about $50$ RNOs in TDRNOT. This is prohibitive with our current 
implementation of solving  the nonlinear EOM \eqref{eom-rno}. We found, for instance, that apart from the expected increase of the numerical effort 
there is the additional complication that  the time step needs to be reduced with increasing $N$.

\begin{figure}[htbp]
\includegraphics[width=0.9\columnwidth]{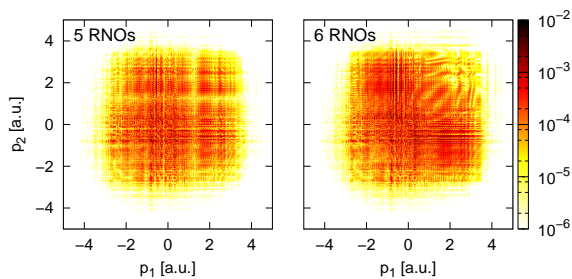}
\caption{(Color online)  $\rho^{2+}(p_1 p_2)$  at $I=2.25\times 10^{15}\; \mathrm{W/cm^2}$ obtained from TDRNOT with $N=15$ spatial RNOs propagated but only $N=5$ and $6$ used to calculate $\rho^{2+}(p_1 p_2)$.}
\label{fig:p2dim_TDROT_}
\end{figure}

Because of the truncation error, the $N$th of the (according ON ordered) $N$ dominant spatial RNOs is expected to be most defective. 
Thus it may make sense to propagate more RNOs than are actually used to calculate observables. Figure~\ref{fig:p2dim_TDROT_} shows results where 
$N=15$ RNOs per spin were propagated but only $N=5$ and $6$ were used for the calculation of the photoelectron spectra. One sees that the agreement 
with the two  corresponding middle-row spectra in Fig.~\ref{fig:p2dim} is much better than in  Fig.~\ref{fig:p2dim_TDROT}. 

%

\subsection{Numerical effort}
The computational time $\tau(N, N_x)$ required for a TDRNOT propagation using $N$ NOs on $N_x$ spatial grid points scales as
\begin{align}
  \tau(N, N_x)
    &\sim
      \alpha\,  N^2\,N_x\,\log N_x + \beta\, N^3\,N_x \label{eq:scaling}
\end{align}
for a fixed time step.
The first term on the right-hand side arises from the calculation of the potentials in \eqref{eom-end} using the fast Fourier transform, and the 
second term from  the evaluation of the required matrix elements in \eqref{eom-rno-B} \cite{remark}. The   computational costs of the corresponding 
operations are taken into account by the constant factors $\alpha$ and $\beta$.

 The computational times  $\tau(N)=\tau(N, 6000)$ required for one laser intensity using $1$ (equivalent to TDHF), $2$, and $6$ NOs were 
$\tau(1)\approx 
1.3\,\mathrm{min}$, $\tau(2)\approx 5.6\,\mathrm{min}$, and $\tau(6)\approx 40\,\mathrm{min}$, respectively, on a single core of an i5-3570 
processor. 
This shows that the $N^2$ term in \eqref{eq:scaling} is dominating.   Compared to the TDSE 
calculation, where  $\tau_\text{TDSE}=12\,\mathrm h$, TDRNOT thus performs faster by a 
factor of $550$, $128$, and  $18$, respectively. 

Unfortunately, the time step in our present TDRNOT implementation needs to be decreased with increasing $N$ to achieve converged results. For 
example, for $N=10$ NOs,  $\Delta t(10) =0.0016$ was used, whereas in the TDSE simulation $\Delta t =0.075$ was sufficient. This leads to a total 
computation time of $\tau(10)=30\,\mathrm h>\tau_\text{TDSE}$ and $\tau(15)=22\,\mathrm d$. Hence, improving our
TDRNOT scheme to allow for larger time steps is desirable.
However, note that for more than two particles the small TDRNOT time step is harmless anyway compared to the  exponential scaling of the TDSE 
wavefunction.

\section{Conclusion and outlook}
In summary, we reproduced the nonsequential double-ionization knee for a He-model atom starting from the spin-singlet ground state using the recently 
introduced time-dependent renormalized-natural-orbital theory. The equations of motion for the  renormalized-natural-orbitals are exact in the 
two-electron case. This is because the expansion of the time-dependent two-body density matrix in  natural orbitals  is known exactly. Only the 
practical limitation in the number of spatial orbitals $N$ forces us to restrict ourselves to $N<20$. Correlated structures in the photoelectron 
spectra are also reproduced. However, a quantitative agreement with the benchmark spectra obtained from the solution of the corresponding two-electron 
time-dependent Schr\"odinger equation  can only be achieved with more orbitals.

Current work is devoted to the application of  time-dependent renormalized-natural-orbital theory to He in full dimensionality, to more electrons, and 
to the mitigation of the truncation error via better-suited boundary conditions.

\section*{Acknowledgment}
This work was supported by the SFB 652 of the German Science Foundation (DFG).


\begin{thebibliography}{38}%
\makeatletter
\providecommand \@ifxundefined [1]{%
 \@ifx{#1\undefined}
}%
\providecommand \@ifnum [1]{%
 \ifnum #1\expandafter \@firstoftwo
 \else \expandafter \@secondoftwo
 \fi
}%
\providecommand \@ifx [1]{%
 \ifx #1\expandafter \@firstoftwo
 \else \expandafter \@secondoftwo
 \fi
}%
\providecommand \natexlab [1]{#1}%
\providecommand \enquote  [1]{``#1''}%
\providecommand \bibnamefont  [1]{#1}%
\providecommand \bibfnamefont [1]{#1}%
\providecommand \citenamefont [1]{#1}%
\providecommand \href@noop [0]{\@secondoftwo}%
\providecommand \href [0]{\begingroup \@sanitize@url \@href}%
\providecommand \@href[1]{\@@startlink{#1}\@@href}%
\providecommand \@@href[1]{\endgroup#1\@@endlink}%
\providecommand \@sanitize@url [0]{\catcode `\\12\catcode `\$12\catcode
  `\&12\catcode `\#12\catcode `\^12\catcode `\_12\catcode `\%12\relax}%
\providecommand \@@startlink[1]{}%
\providecommand \@@endlink[0]{}%
\providecommand \url  [0]{\begingroup\@sanitize@url \@url }%
\providecommand \@url [1]{\endgroup\@href {#1}{\urlprefix }}%
\providecommand \urlprefix  [0]{URL }%
\providecommand \Eprint [0]{\href }%
\providecommand \doibase [0]{http://dx.doi.org/}%
\providecommand \selectlanguage [0]{\@gobble}%
\providecommand \bibinfo  [0]{\@secondoftwo}%
\providecommand \bibfield  [0]{\@secondoftwo}%
\providecommand \translation [1]{[#1]}%
\providecommand \BibitemOpen [0]{}%
\providecommand \bibitemStop [0]{}%
\providecommand \bibitemNoStop [0]{.\EOS\space}%
\providecommand \EOS [0]{\spacefactor3000\relax}%
\providecommand \BibitemShut  [1]{\csname bibitem#1\endcsname}%
\let\auto@bib@innerbib\@empty
\bibitem [{\citenamefont {Becker}\ \emph {et~al.}(2012)\citenamefont {Becker},
  \citenamefont {Liu}, \citenamefont {Ho},\ and\ \citenamefont
  {Eberly}}]{WBeckerRev}%
  \BibitemOpen
  \bibfield  {author} {\bibinfo {author} {\bibfnamefont {W.}~\bibnamefont
  {Becker}}, \bibinfo {author} {\bibfnamefont {X.}~\bibnamefont {Liu}},
  \bibinfo {author} {\bibfnamefont {P.~J.}\ \bibnamefont {Ho}}, \ and\ \bibinfo
  {author} {\bibfnamefont {J.~H.}\ \bibnamefont {Eberly}},\ }\href {\doibase
  10.1103/RevModPhys.84.1011} {\bibfield  {journal} {\bibinfo  {journal} {Rev.
  Mod. Phys.}\ }\textbf {\bibinfo {volume} {84}},\ \bibinfo {pages} {1011}
  (\bibinfo {year} {2012})}\BibitemShut {NoStop}%
\bibitem [{\citenamefont {de~Morisson~Faria}\ and\ \citenamefont
  {Liu}(2011)}]{Carla2011}%
  \BibitemOpen
  \bibfield  {author} {\bibinfo {author} {\bibfnamefont {C.~F.}\ \bibnamefont
  {de~Morisson~Faria}}\ and\ \bibinfo {author} {\bibfnamefont {X.}~\bibnamefont
  {Liu}},\ }\href {\doibase 10.1080/09500340.2010.543958} {\bibfield  {journal}
  {\bibinfo  {journal} {Journal of Modern Optics}\ }\textbf {\bibinfo {volume}
  {58}},\ \bibinfo {pages} {1076} (\bibinfo {year} {2011})}\BibitemShut
  {NoStop}%
\bibitem [{\citenamefont {Weber}\ \emph
  {et~al.}(2000{\natexlab{a}})\citenamefont {Weber}, \citenamefont
  {Weckenbrock}, \citenamefont {Staudte}, \citenamefont {Spielberger},
  \citenamefont {Jagutzki}, \citenamefont {Mergel}, \citenamefont {Afaneh},
  \citenamefont {Urbasch}, \citenamefont {Vollmer}, \citenamefont {Giessen},\
  and\ \citenamefont {D\"orner}}]{weberNSDI2000}%
  \BibitemOpen
  \bibfield  {author} {\bibinfo {author} {\bibfnamefont {T.}~\bibnamefont
  {Weber}}, \bibinfo {author} {\bibfnamefont {M.}~\bibnamefont {Weckenbrock}},
  \bibinfo {author} {\bibfnamefont {A.}~\bibnamefont {Staudte}}, \bibinfo
  {author} {\bibfnamefont {L.}~\bibnamefont {Spielberger}}, \bibinfo {author}
  {\bibfnamefont {O.}~\bibnamefont {Jagutzki}}, \bibinfo {author}
  {\bibfnamefont {V.}~\bibnamefont {Mergel}}, \bibinfo {author} {\bibfnamefont
  {F.}~\bibnamefont {Afaneh}}, \bibinfo {author} {\bibfnamefont
  {G.}~\bibnamefont {Urbasch}}, \bibinfo {author} {\bibfnamefont
  {M.}~\bibnamefont {Vollmer}}, \bibinfo {author} {\bibfnamefont
  {H.}~\bibnamefont {Giessen}}, \ and\ \bibinfo {author} {\bibfnamefont
  {R.}~\bibnamefont {D\"orner}},\ }\href {\doibase 10.1103/PhysRevLett.84.443}
  {\bibfield  {journal} {\bibinfo  {journal} {Phys. Rev. Lett.}\ }\textbf
  {\bibinfo {volume} {84}},\ \bibinfo {pages} {443} (\bibinfo {year}
  {2000}{\natexlab{a}})}\BibitemShut {NoStop}%
\bibitem [{\citenamefont {Moshammer}\ \emph {et~al.}(2000)\citenamefont
  {Moshammer}, \citenamefont {Feuerstein}, \citenamefont {Schmitt},
  \citenamefont {Dorn}, \citenamefont {Schr\"oter}, \citenamefont {Ullrich},
  \citenamefont {Rottke}, \citenamefont {Trump}, \citenamefont {Wittmann},
  \citenamefont {Korn}, \citenamefont {Hoffmann},\ and\ \citenamefont
  {Sandner}}]{moshammerNSDI2000}%
  \BibitemOpen
  \bibfield  {author} {\bibinfo {author} {\bibfnamefont {R.}~\bibnamefont
  {Moshammer}}, \bibinfo {author} {\bibfnamefont {B.}~\bibnamefont
  {Feuerstein}}, \bibinfo {author} {\bibfnamefont {W.}~\bibnamefont {Schmitt}},
  \bibinfo {author} {\bibfnamefont {A.}~\bibnamefont {Dorn}}, \bibinfo {author}
  {\bibfnamefont {C.~D.}\ \bibnamefont {Schr\"oter}}, \bibinfo {author}
  {\bibfnamefont {J.}~\bibnamefont {Ullrich}}, \bibinfo {author} {\bibfnamefont
  {H.}~\bibnamefont {Rottke}}, \bibinfo {author} {\bibfnamefont
  {C.}~\bibnamefont {Trump}}, \bibinfo {author} {\bibfnamefont
  {M.}~\bibnamefont {Wittmann}}, \bibinfo {author} {\bibfnamefont
  {G.}~\bibnamefont {Korn}}, \bibinfo {author} {\bibfnamefont {K.}~\bibnamefont
  {Hoffmann}}, \ and\ \bibinfo {author} {\bibfnamefont {W.}~\bibnamefont
  {Sandner}},\ }\href {\doibase 10.1103/PhysRevLett.84.447} {\bibfield
  {journal} {\bibinfo  {journal} {Phys. Rev. Lett.}\ }\textbf {\bibinfo
  {volume} {84}},\ \bibinfo {pages} {447} (\bibinfo {year} {2000})}\BibitemShut
  {NoStop}%
\bibitem [{\citenamefont {Weber}\ \emph
  {et~al.}(2000{\natexlab{b}})\citenamefont {Weber}, \citenamefont {Giessen},
  \citenamefont {Weckenbrock}, \citenamefont {Urbasch}, \citenamefont
  {Staudte}, \citenamefont {Spielberger}, \citenamefont {Jagutzki},
  \citenamefont {Mergel}, \citenamefont {Vollmer},\ and\ \citenamefont
  {D\"orner}}]{webernature2000}%
  \BibitemOpen
  \bibfield  {author} {\bibinfo {author} {\bibfnamefont {T.}~\bibnamefont
  {Weber}}, \bibinfo {author} {\bibfnamefont {H.}~\bibnamefont {Giessen}},
  \bibinfo {author} {\bibfnamefont {M.}~\bibnamefont {Weckenbrock}}, \bibinfo
  {author} {\bibfnamefont {G.}~\bibnamefont {Urbasch}}, \bibinfo {author}
  {\bibfnamefont {A.}~\bibnamefont {Staudte}}, \bibinfo {author} {\bibfnamefont
  {L.}~\bibnamefont {Spielberger}}, \bibinfo {author} {\bibfnamefont
  {O.}~\bibnamefont {Jagutzki}}, \bibinfo {author} {\bibfnamefont
  {V.}~\bibnamefont {Mergel}}, \bibinfo {author} {\bibfnamefont
  {M.}~\bibnamefont {Vollmer}}, \ and\ \bibinfo {author} {\bibfnamefont
  {R.}~\bibnamefont {D\"orner}},\ }\href@noop {} {\bibfield  {journal}
  {\bibinfo  {journal} {Nature}\ }\textbf {\bibinfo {volume} {405}},\ \bibinfo
  {pages} {658} (\bibinfo {year} {2000}{\natexlab{b}})}\BibitemShut {NoStop}%
\bibitem [{\citenamefont {Feuerstein}\ \emph {et~al.}(2001)\citenamefont
  {Feuerstein}, \citenamefont {Moshammer}, \citenamefont {Fischer},
  \citenamefont {Dorn}, \citenamefont {Schr\"oter}, \citenamefont
  {Deipenwisch}, \citenamefont {Crespo Lopez-Urrutia}, \citenamefont {H\"ohr},
  \citenamefont {Neumayer}, \citenamefont {Ullrich}, \citenamefont {Rottke},
  \citenamefont {Trump}, \citenamefont {Wittmann}, \citenamefont {Korn},\ and\
  \citenamefont {Sandner}}]{feuerstein2001}%
  \BibitemOpen
  \bibfield  {author} {\bibinfo {author} {\bibfnamefont {B.}~\bibnamefont
  {Feuerstein}}, \bibinfo {author} {\bibfnamefont {R.}~\bibnamefont
  {Moshammer}}, \bibinfo {author} {\bibfnamefont {D.}~\bibnamefont {Fischer}},
  \bibinfo {author} {\bibfnamefont {A.}~\bibnamefont {Dorn}}, \bibinfo {author}
  {\bibfnamefont {C.~D.}\ \bibnamefont {Schr\"oter}}, \bibinfo {author}
  {\bibfnamefont {J.}~\bibnamefont {Deipenwisch}}, \bibinfo {author}
  {\bibfnamefont {J.~R.}\ \bibnamefont {Crespo Lopez-Urrutia}}, \bibinfo
  {author} {\bibfnamefont {C.}~\bibnamefont {H\"ohr}}, \bibinfo {author}
  {\bibfnamefont {P.}~\bibnamefont {Neumayer}}, \bibinfo {author}
  {\bibfnamefont {J.}~\bibnamefont {Ullrich}}, \bibinfo {author} {\bibfnamefont
  {H.}~\bibnamefont {Rottke}}, \bibinfo {author} {\bibfnamefont
  {C.}~\bibnamefont {Trump}}, \bibinfo {author} {\bibfnamefont
  {M.}~\bibnamefont {Wittmann}}, \bibinfo {author} {\bibfnamefont
  {G.}~\bibnamefont {Korn}}, \ and\ \bibinfo {author} {\bibfnamefont
  {W.}~\bibnamefont {Sandner}},\ }\href {\doibase
  10.1103/PhysRevLett.87.043003} {\bibfield  {journal} {\bibinfo  {journal}
  {Phys. Rev. Lett.}\ }\textbf {\bibinfo {volume} {87}},\ \bibinfo {pages}
  {043003} (\bibinfo {year} {2001})}\BibitemShut {NoStop}%
\bibitem [{\citenamefont {Kübel}\ \emph {et~al.}(2014)\citenamefont {Kübel},
  \citenamefont {Betsch}, \citenamefont {Kling}, \citenamefont {Alnaser},
  \citenamefont {Schmidt}, \citenamefont {Kleineberg}, \citenamefont {Deng},
  \citenamefont {Ben-Itzhak}, \citenamefont {Paulus}, \citenamefont {Pfeifer},
  \citenamefont {Ullrich}, \citenamefont {Moshammer}, \citenamefont {Kling},\
  and\ \citenamefont {Bergues}}]{kuebel}%
  \BibitemOpen
  \bibfield  {author} {\bibinfo {author} {\bibfnamefont {M.}~\bibnamefont
  {Kübel}}, \bibinfo {author} {\bibfnamefont {K.~J.}\ \bibnamefont {Betsch}},
  \bibinfo {author} {\bibfnamefont {N.~G.}\ \bibnamefont {Kling}}, \bibinfo
  {author} {\bibfnamefont {A.~S.}\ \bibnamefont {Alnaser}}, \bibinfo {author}
  {\bibfnamefont {J.}~\bibnamefont {Schmidt}}, \bibinfo {author} {\bibfnamefont
  {U.}~\bibnamefont {Kleineberg}}, \bibinfo {author} {\bibfnamefont
  {Y.}~\bibnamefont {Deng}}, \bibinfo {author} {\bibfnamefont {I.}~\bibnamefont
  {Ben-Itzhak}}, \bibinfo {author} {\bibfnamefont {G.~G.}\ \bibnamefont
  {Paulus}}, \bibinfo {author} {\bibfnamefont {T.}~\bibnamefont {Pfeifer}},
  \bibinfo {author} {\bibfnamefont {J.}~\bibnamefont {Ullrich}}, \bibinfo
  {author} {\bibfnamefont {R.}~\bibnamefont {Moshammer}}, \bibinfo {author}
  {\bibfnamefont {M.~F.}\ \bibnamefont {Kling}}, \ and\ \bibinfo {author}
  {\bibfnamefont {B.}~\bibnamefont {Bergues}},\ }\href
  {http://stacks.iop.org/1367-2630/16/i=3/a=033008} {\bibfield  {journal}
  {\bibinfo  {journal} {New Journal of Physics}\ }\textbf {\bibinfo {volume}
  {16}},\ \bibinfo {pages} {033008} (\bibinfo {year} {2014})}\BibitemShut
  {NoStop}%
\bibitem [{\citenamefont {Rudenko}\ \emph {et~al.}(2004)\citenamefont
  {Rudenko}, \citenamefont {Zrost}, \citenamefont {Feuerstein}, \citenamefont
  {de~Jesus}, \citenamefont {Schr\"oter}, \citenamefont {Moshammer},\ and\
  \citenamefont {Ullrich}}]{RESI}%
  \BibitemOpen
  \bibfield  {author} {\bibinfo {author} {\bibfnamefont {A.}~\bibnamefont
  {Rudenko}}, \bibinfo {author} {\bibfnamefont {K.}~\bibnamefont {Zrost}},
  \bibinfo {author} {\bibfnamefont {B.}~\bibnamefont {Feuerstein}}, \bibinfo
  {author} {\bibfnamefont {V.~L.~B.}\ \bibnamefont {de~Jesus}}, \bibinfo
  {author} {\bibfnamefont {C.~D.}\ \bibnamefont {Schr\"oter}}, \bibinfo
  {author} {\bibfnamefont {R.}~\bibnamefont {Moshammer}}, \ and\ \bibinfo
  {author} {\bibfnamefont {J.}~\bibnamefont {Ullrich}},\ }\href {\doibase
  10.1103/PhysRevLett.93.253001} {\bibfield  {journal} {\bibinfo  {journal}
  {Phys. Rev. Lett.}\ }\textbf {\bibinfo {volume} {93}},\ \bibinfo {pages}
  {253001} (\bibinfo {year} {2004})}\BibitemShut {NoStop}%
\bibitem [{\citenamefont {Dahlen}\ and\ \citenamefont {van
  Leeuwen}(2001)}]{Dahlen_Leeuwen}%
  \BibitemOpen
  \bibfield  {author} {\bibinfo {author} {\bibfnamefont {N.~E.}\ \bibnamefont
  {Dahlen}}\ and\ \bibinfo {author} {\bibfnamefont {R.}~\bibnamefont {van
  Leeuwen}},\ }\href {\doibase 10.1103/PhysRevA.64.023405} {\bibfield
  {journal} {\bibinfo  {journal} {Phys. Rev. A}\ }\textbf {\bibinfo {volume}
  {64}},\ \bibinfo {pages} {023405} (\bibinfo {year} {2001})}\BibitemShut
  {NoStop}%
\bibitem [{\citenamefont {Lein}\ and\ \citenamefont
  {K\"ummel}(2005)}]{MLein2005}%
  \BibitemOpen
  \bibfield  {author} {\bibinfo {author} {\bibfnamefont {M.}~\bibnamefont
  {Lein}}\ and\ \bibinfo {author} {\bibfnamefont {S.}~\bibnamefont
  {K\"ummel}},\ }\href {\doibase 10.1103/PhysRevLett.94.143003} {\bibfield
  {journal} {\bibinfo  {journal} {Phys. Rev. Lett.}\ }\textbf {\bibinfo
  {volume} {94}},\ \bibinfo {pages} {143003} (\bibinfo {year}
  {2005})}\BibitemShut {NoStop}%
\bibitem [{\citenamefont {Petersilka}\ and\ \citenamefont
  {Gross}(1999)}]{petersilka}%
  \BibitemOpen
  \bibfield  {author} {\bibinfo {author} {\bibfnamefont {M.}~\bibnamefont
  {Petersilka}}\ and\ \bibinfo {author} {\bibfnamefont {E.~K.~U.}\ \bibnamefont
  {Gross}},\ }\href@noop {} {\bibfield  {journal} {\bibinfo  {journal} {Laser
  Phys.}\ }\textbf {\bibinfo {volume} {9}},\ \bibinfo {pages} {105} (\bibinfo
  {year} {1999})}\BibitemShut {NoStop}%
\bibitem [{\citenamefont {Wilken}\ and\ \citenamefont {Bauer}(2006)}]{florian}%
  \BibitemOpen
  \bibfield  {author} {\bibinfo {author} {\bibfnamefont {F.}~\bibnamefont
  {Wilken}}\ and\ \bibinfo {author} {\bibfnamefont {D.}~\bibnamefont {Bauer}},\
  }\href {\doibase 10.1103/PhysRevLett.97.203001} {\bibfield  {journal}
  {\bibinfo  {journal} {Phys. Rev. Lett.}\ }\textbf {\bibinfo {volume} {97}},\
  \bibinfo {pages} {203001} (\bibinfo {year} {2006})}\BibitemShut {NoStop}%
\bibitem [{\citenamefont {Parker}\ \emph {et~al.}(2006)\citenamefont {Parker},
  \citenamefont {Doherty}, \citenamefont {Taylor}, \citenamefont {Schultz},
  \citenamefont {Blaga},\ and\ \citenamefont {DiMauro}}]{KenTaylor2006}%
  \BibitemOpen
  \bibfield  {author} {\bibinfo {author} {\bibfnamefont {J.~S.}\ \bibnamefont
  {Parker}}, \bibinfo {author} {\bibfnamefont {B.~J.~S.}\ \bibnamefont
  {Doherty}}, \bibinfo {author} {\bibfnamefont {K.~T.}\ \bibnamefont {Taylor}},
  \bibinfo {author} {\bibfnamefont {K.~D.}\ \bibnamefont {Schultz}}, \bibinfo
  {author} {\bibfnamefont {C.~I.}\ \bibnamefont {Blaga}}, \ and\ \bibinfo
  {author} {\bibfnamefont {L.~F.}\ \bibnamefont {DiMauro}},\ }\href {\doibase
  10.1103/PhysRevLett.96.133001} {\bibfield  {journal} {\bibinfo  {journal}
  {Phys. Rev. Lett.}\ }\textbf {\bibinfo {volume} {96}},\ \bibinfo {pages}
  {133001} (\bibinfo {year} {2006})}\BibitemShut {NoStop}%
\bibitem [{\citenamefont {Rohringer}\ \emph {et~al.}(2006)\citenamefont
  {Rohringer}, \citenamefont {Gordon},\ and\ \citenamefont
  {Santra}}]{Rohringer}%
  \BibitemOpen
  \bibfield  {author} {\bibinfo {author} {\bibfnamefont {N.}~\bibnamefont
  {Rohringer}}, \bibinfo {author} {\bibfnamefont {A.}~\bibnamefont {Gordon}}, \
  and\ \bibinfo {author} {\bibfnamefont {R.}~\bibnamefont {Santra}},\ }\href
  {\doibase 10.1103/PhysRevA.74.043420} {\bibfield  {journal} {\bibinfo
  {journal} {Phys. Rev. A}\ }\textbf {\bibinfo {volume} {74}},\ \bibinfo
  {pages} {043420} (\bibinfo {year} {2006})}\BibitemShut {NoStop}%
\bibitem [{\citenamefont {Karamatskou}\ \emph {et~al.}(2014)\citenamefont
  {Karamatskou}, \citenamefont {Pabst}, \citenamefont {Chen},\ and\
  \citenamefont {Santra}}]{Santra}%
  \BibitemOpen
  \bibfield  {author} {\bibinfo {author} {\bibfnamefont {A.}~\bibnamefont
  {Karamatskou}}, \bibinfo {author} {\bibfnamefont {S.}~\bibnamefont {Pabst}},
  \bibinfo {author} {\bibfnamefont {Y.-J.}\ \bibnamefont {Chen}}, \ and\
  \bibinfo {author} {\bibfnamefont {R.}~\bibnamefont {Santra}},\ }\href
  {\doibase 10.1103/PhysRevA.89.033415} {\bibfield  {journal} {\bibinfo
  {journal} {Phys. Rev. A}\ }\textbf {\bibinfo {volume} {89}},\ \bibinfo
  {pages} {033415} (\bibinfo {year} {2014})}\BibitemShut {NoStop}%
\bibitem [{\citenamefont {Sukiasyan}\ \emph {et~al.}(2009)\citenamefont
  {Sukiasyan}, \citenamefont {McDonald}, \citenamefont {Van~Vlack},
  \citenamefont {Destefani}, \citenamefont {Fennel}, \citenamefont {Ivanov},\
  and\ \citenamefont {Brabec}}]{mctdh}%
  \BibitemOpen
  \bibfield  {author} {\bibinfo {author} {\bibfnamefont {S.}~\bibnamefont
  {Sukiasyan}}, \bibinfo {author} {\bibfnamefont {C.}~\bibnamefont {McDonald}},
  \bibinfo {author} {\bibfnamefont {C.}~\bibnamefont {Van~Vlack}}, \bibinfo
  {author} {\bibfnamefont {C.}~\bibnamefont {Destefani}}, \bibinfo {author}
  {\bibfnamefont {T.}~\bibnamefont {Fennel}}, \bibinfo {author} {\bibfnamefont
  {M.}~\bibnamefont {Ivanov}}, \ and\ \bibinfo {author} {\bibfnamefont
  {T.}~\bibnamefont {Brabec}},\ }\href {\doibase 10.1103/PhysRevA.80.013412}
  {\bibfield  {journal} {\bibinfo  {journal} {Phys. Rev. A}\ }\textbf {\bibinfo
  {volume} {80}},\ \bibinfo {pages} {013412} (\bibinfo {year}
  {2009})}\BibitemShut {NoStop}%
\bibitem [{\citenamefont {Zanghellini}\ \emph {et~al.}(2004)\citenamefont
  {Zanghellini}, \citenamefont {Kitzler}, \citenamefont {Brabec},\ and\
  \citenamefont {Scrinzi}}]{Zanghellini}%
  \BibitemOpen
  \bibfield  {author} {\bibinfo {author} {\bibfnamefont {J.}~\bibnamefont
  {Zanghellini}}, \bibinfo {author} {\bibfnamefont {M.}~\bibnamefont
  {Kitzler}}, \bibinfo {author} {\bibfnamefont {T.}~\bibnamefont {Brabec}}, \
  and\ \bibinfo {author} {\bibfnamefont {A.}~\bibnamefont {Scrinzi}},\ }\href
  {http://stacks.iop.org/0953-4075/37/i=4/a=004} {\bibfield  {journal}
  {\bibinfo  {journal} {Journal of Physics B: Atomic, Molecular and Optical
  Physics}\ }\textbf {\bibinfo {volume} {37}},\ \bibinfo {pages} {763}
  (\bibinfo {year} {2004})}\BibitemShut {NoStop}%
\bibitem [{\citenamefont {Caillat}\ \emph {et~al.}(2005)\citenamefont
  {Caillat}, \citenamefont {Zanghellini}, \citenamefont {Kitzler},
  \citenamefont {Koch}, \citenamefont {Kreuzer},\ and\ \citenamefont
  {Scrinzi}}]{caillat}%
  \BibitemOpen
  \bibfield  {author} {\bibinfo {author} {\bibfnamefont {J.}~\bibnamefont
  {Caillat}}, \bibinfo {author} {\bibfnamefont {J.}~\bibnamefont
  {Zanghellini}}, \bibinfo {author} {\bibfnamefont {M.}~\bibnamefont
  {Kitzler}}, \bibinfo {author} {\bibfnamefont {O.}~\bibnamefont {Koch}},
  \bibinfo {author} {\bibfnamefont {W.}~\bibnamefont {Kreuzer}}, \ and\
  \bibinfo {author} {\bibfnamefont {A.}~\bibnamefont {Scrinzi}},\ }\href
  {\doibase 10.1103/PhysRevA.71.012712} {\bibfield  {journal} {\bibinfo
  {journal} {Phys. Rev. A}\ }\textbf {\bibinfo {volume} {71}},\ \bibinfo
  {pages} {012712} (\bibinfo {year} {2005})}\BibitemShut {NoStop}%
\bibitem [{\citenamefont {Koch}\ \emph {et~al.}(2006)\citenamefont {Koch},
  \citenamefont {Kreuzer},\ and\ \citenamefont {Scrinzi}}]{Scrinzi2006}%
  \BibitemOpen
  \bibfield  {author} {\bibinfo {author} {\bibfnamefont {O.}~\bibnamefont
  {Koch}}, \bibinfo {author} {\bibfnamefont {W.}~\bibnamefont {Kreuzer}}, \
  and\ \bibinfo {author} {\bibfnamefont {A.}~\bibnamefont {Scrinzi}},\ }\href
  {\doibase http://dx.doi.org/10.1016/j.amc.2005.04.027} {\bibfield  {journal}
  {\bibinfo  {journal} {Applied Mathematics and Computation}\ }\textbf
  {\bibinfo {volume} {173}},\ \bibinfo {pages} {960 } (\bibinfo {year}
  {2006})}\BibitemShut {NoStop}%
\bibitem [{\citenamefont {Hochstuhl}\ \emph {et~al.}(2010)\citenamefont
  {Hochstuhl}, \citenamefont {Bauch},\ and\ \citenamefont {Bonitz}}]{mctdhf}%
  \BibitemOpen
  \bibfield  {author} {\bibinfo {author} {\bibfnamefont {D.}~\bibnamefont
  {Hochstuhl}}, \bibinfo {author} {\bibfnamefont {S.}~\bibnamefont {Bauch}}, \
  and\ \bibinfo {author} {\bibfnamefont {M.}~\bibnamefont {Bonitz}},\ }\href
  {http://stacks.iop.org/1742-6596/220/i=1/a=012019} {\bibfield  {journal}
  {\bibinfo  {journal} {Journal of Physics: Conference Series}\ }\textbf
  {\bibinfo {volume} {220}},\ \bibinfo {pages} {012019} (\bibinfo {year}
  {2010})}\BibitemShut {NoStop}%
\bibitem [{\citenamefont {Nguyen}\ and\ \citenamefont
  {Bandrauk}(2006)}]{Bandrauk2006}%
  \BibitemOpen
  \bibfield  {author} {\bibinfo {author} {\bibfnamefont {N.~A.}\ \bibnamefont
  {Nguyen}}\ and\ \bibinfo {author} {\bibfnamefont {A.~D.}\ \bibnamefont
  {Bandrauk}},\ }\href {\doibase 10.1103/PhysRevA.73.032708} {\bibfield
  {journal} {\bibinfo  {journal} {Phys. Rev. A}\ }\textbf {\bibinfo {volume}
  {73}},\ \bibinfo {pages} {032708} (\bibinfo {year} {2006})}\BibitemShut
  {NoStop}%
\bibitem [{\citenamefont {Watson}\ \emph {et~al.}(1997)\citenamefont {Watson},
  \citenamefont {Sanpera}, \citenamefont {Lappas}, \citenamefont {Knight},\
  and\ \citenamefont {Burnett}}]{Watson1997}%
  \BibitemOpen
  \bibfield  {author} {\bibinfo {author} {\bibfnamefont {J.~B.}\ \bibnamefont
  {Watson}}, \bibinfo {author} {\bibfnamefont {A.}~\bibnamefont {Sanpera}},
  \bibinfo {author} {\bibfnamefont {D.~G.}\ \bibnamefont {Lappas}}, \bibinfo
  {author} {\bibfnamefont {P.~L.}\ \bibnamefont {Knight}}, \ and\ \bibinfo
  {author} {\bibfnamefont {K.}~\bibnamefont {Burnett}},\ }\href {\doibase
  10.1103/PhysRevLett.78.1884} {\bibfield  {journal} {\bibinfo  {journal}
  {Phys. Rev. Lett.}\ }\textbf {\bibinfo {volume} {78}},\ \bibinfo {pages}
  {1884} (\bibinfo {year} {1997})}\BibitemShut {NoStop}%
\bibitem [{\citenamefont {Kohn}(1999)}]{KohnNobel1999}%
  \BibitemOpen
  \bibfield  {author} {\bibinfo {author} {\bibfnamefont {W.}~\bibnamefont
  {Kohn}},\ }\href {\doibase 10.1103/RevModPhys.71.1253} {\bibfield  {journal}
  {\bibinfo  {journal} {Rev. Mod. Phys.}\ }\textbf {\bibinfo {volume} {71}},\
  \bibinfo {pages} {1253} (\bibinfo {year} {1999})}\BibitemShut {NoStop}%
\bibitem [{\citenamefont {Brics}\ and\ \citenamefont {Bauer}(2013)}]{tdrnot}%
  \BibitemOpen
  \bibfield  {author} {\bibinfo {author} {\bibfnamefont {M.}~\bibnamefont
  {Brics}}\ and\ \bibinfo {author} {\bibfnamefont {D.}~\bibnamefont {Bauer}},\
  }\href {\doibase 10.1103/PhysRevA.88.052514} {\bibfield  {journal} {\bibinfo
  {journal} {Phys. Rev. A}\ }\textbf {\bibinfo {volume} {88}},\ \bibinfo
  {pages} {052514} (\bibinfo {year} {2013})}\BibitemShut {NoStop}%
\bibitem [{\citenamefont {Rapp}\ \emph {et~al.}(2014)\citenamefont {Rapp},
  \citenamefont {Brics},\ and\ \citenamefont {Bauer}}]{tdrnot2}%
  \BibitemOpen
  \bibfield  {author} {\bibinfo {author} {\bibfnamefont {J.}~\bibnamefont
  {Rapp}}, \bibinfo {author} {\bibfnamefont {M.}~\bibnamefont {Brics}}, \ and\
  \bibinfo {author} {\bibfnamefont {D.}~\bibnamefont {Bauer}},\ }\href
  {\doibase 10.1103/PhysRevA.90.012518} {\bibfield  {journal} {\bibinfo
  {journal} {Phys. Rev. A}\ }\textbf {\bibinfo {volume} {90}},\ \bibinfo
  {pages} {012518} (\bibinfo {year} {2014})}\BibitemShut {NoStop}%
\bibitem [{\citenamefont {Bauer}(1997)}]{Dieter97}%
  \BibitemOpen
  \bibfield  {author} {\bibinfo {author} {\bibfnamefont {D.}~\bibnamefont
  {Bauer}},\ }\href {\doibase 10.1103/PhysRevA.56.3028} {\bibfield  {journal}
  {\bibinfo  {journal} {Phys. Rev. A}\ }\textbf {\bibinfo {volume} {56}},\
  \bibinfo {pages} {3028} (\bibinfo {year} {1997})}\BibitemShut {NoStop}%
\bibitem [{\citenamefont {Lappas}\ and\ \citenamefont {van
  Leeuwen}(1998)}]{Lappas_Leeuwen}%
  \BibitemOpen
  \bibfield  {author} {\bibinfo {author} {\bibfnamefont {D.~G.}\ \bibnamefont
  {Lappas}}\ and\ \bibinfo {author} {\bibfnamefont {R.}~\bibnamefont {van
  Leeuwen}},\ }\href {\doibase 10.1088/0953-4075/31/6/001} {\bibfield
  {journal} {\bibinfo  {journal} {J. Phys. B}\ }\textbf {\bibinfo {volume}
  {31}},\ \bibinfo {pages} {L249} (\bibinfo {year} {1998})}\BibitemShut
  {NoStop}%
\bibitem [{\citenamefont {Lein}\ \emph {et~al.}(2000)\citenamefont {Lein},
  \citenamefont {Gross},\ and\ \citenamefont {Engel}}]{Lein2000}%
  \BibitemOpen
  \bibfield  {author} {\bibinfo {author} {\bibfnamefont {M.}~\bibnamefont
  {Lein}}, \bibinfo {author} {\bibfnamefont {E.~K.~U.}\ \bibnamefont {Gross}},
  \ and\ \bibinfo {author} {\bibfnamefont {V.}~\bibnamefont {Engel}},\ }\href
  {\doibase 10.1103/PhysRevLett.85.4707} {\bibfield  {journal} {\bibinfo
  {journal} {Phys. Rev. Lett.}\ }\textbf {\bibinfo {volume} {85}},\ \bibinfo
  {pages} {4707} (\bibinfo {year} {2000})}\BibitemShut {NoStop}%
\bibitem [{\citenamefont {Wilken}\ and\ \citenamefont
  {Bauer}(2007)}]{florian2}%
  \BibitemOpen
  \bibfield  {author} {\bibinfo {author} {\bibfnamefont {F.}~\bibnamefont
  {Wilken}}\ and\ \bibinfo {author} {\bibfnamefont {D.}~\bibnamefont {Bauer}},\
  }\href {\doibase 10.1103/PhysRevA.76.023409} {\bibfield  {journal} {\bibinfo
  {journal} {Phys. Rev. A}\ }\textbf {\bibinfo {volume} {76}},\ \bibinfo
  {pages} {023409} (\bibinfo {year} {2007})}\BibitemShut {NoStop}%
\bibitem [{\citenamefont {L\"owdin}(1955)}]{loewdin}%
  \BibitemOpen
  \bibfield  {author} {\bibinfo {author} {\bibfnamefont {P.-O.}\ \bibnamefont
  {L\"owdin}},\ }\href {\doibase 10.1103/PhysRev.97.1474} {\bibfield  {journal}
  {\bibinfo  {journal} {Phys. Rev.}\ }\textbf {\bibinfo {volume} {97}},\
  \bibinfo {pages} {1474} (\bibinfo {year} {1955})}\BibitemShut {NoStop}%
\bibitem [{\citenamefont {Coleman}(1963)}]{coleman}%
  \BibitemOpen
  \bibfield  {author} {\bibinfo {author} {\bibfnamefont {A.~J.}\ \bibnamefont
  {Coleman}},\ }\href {\doibase 10.1103/RevModPhys.35.668} {\bibfield
  {journal} {\bibinfo  {journal} {Rev. Mod. Phys.}\ }\textbf {\bibinfo {volume}
  {35}},\ \bibinfo {pages} {668} (\bibinfo {year} {1963})}\BibitemShut
  {NoStop}%
\bibitem [{\citenamefont {Coleman}\ and\ \citenamefont
  {Yukalov}(2000)}]{colemanorangebook}%
  \BibitemOpen
  \bibfield  {author} {\bibinfo {author} {\bibfnamefont {A.}~\bibnamefont
  {Coleman}}\ and\ \bibinfo {author} {\bibfnamefont {V.}~\bibnamefont
  {Yukalov}},\ }\href@noop {} {\emph {\bibinfo {title} {Reduced Density
  Matrices, Coulson's Challenge}}},\ Springer Lecture Notes in Chemistry 72\
  (\bibinfo  {publisher} {Springer},\ \bibinfo {address} {Berlin},\ \bibinfo
  {year} {2000})\BibitemShut {NoStop}%
\bibitem [{\citenamefont {Pernal}\ \emph {et~al.}(2007)\citenamefont {Pernal},
  \citenamefont {Gritsenko},\ and\ \citenamefont {Baerends}}]{Pernal}%
  \BibitemOpen
  \bibfield  {author} {\bibinfo {author} {\bibfnamefont {K.}~\bibnamefont
  {Pernal}}, \bibinfo {author} {\bibfnamefont {O.}~\bibnamefont {Gritsenko}}, \
  and\ \bibinfo {author} {\bibfnamefont {E.~J.}\ \bibnamefont {Baerends}},\
  }\href {\doibase 10.1103/PhysRevA.75.012506} {\bibfield  {journal} {\bibinfo
  {journal} {Phys. Rev. A}\ }\textbf {\bibinfo {volume} {75}},\ \bibinfo
  {pages} {012506} (\bibinfo {year} {2007})}\BibitemShut {NoStop}%
\bibitem [{\citenamefont {Appel}(2007)}]{appel}%
  \BibitemOpen
  \bibfield  {author} {\bibinfo {author} {\bibfnamefont {H.}~\bibnamefont
  {Appel}},\ }\emph {\bibinfo {title} {Time-Dependent Quantum Many-Body
  Systems: Linear Response, Electronic Transport, and Reduced Density
  Matrices}},\ \href@noop {} {Ph.D. thesis},\ \bibinfo  {school} {Freie
  Universität Berlin} (\bibinfo {year} {2007})\BibitemShut {NoStop}%
\bibitem [{\citenamefont {Giesbertz}(2010)}]{giesbertz}%
  \BibitemOpen
  \bibfield  {author} {\bibinfo {author} {\bibfnamefont {K.~J.~H.}\
  \bibnamefont {Giesbertz}},\ }\emph {\bibinfo {title} {Time-Dependent One-Body
  Reduced Density Matrix Functional Theory}},\ \href@noop {} {Ph.D. thesis},\
  \bibinfo  {school} {Free University Amsterdam} (\bibinfo {year}
  {2010})\BibitemShut {NoStop}%
\bibitem [{\citenamefont {Appel}\ and\ \citenamefont
  {Gross}(2010)}]{appelgross}%
  \BibitemOpen
  \bibfield  {author} {\bibinfo {author} {\bibfnamefont {H.}~\bibnamefont
  {Appel}}\ and\ \bibinfo {author} {\bibfnamefont {E.~K.~U.}\ \bibnamefont
  {Gross}},\ }\href {http://stacks.iop.org/0295-5075/92/i=2/a=23001} {\bibfield
   {journal} {\bibinfo  {journal} {EPL (Europhysics Letters)}\ }\textbf
  {\bibinfo {volume} {92}},\ \bibinfo {pages} {23001} (\bibinfo {year}
  {2010})}\BibitemShut {NoStop}%
\bibitem [{\citenamefont {Helbig}\ \emph {et~al.}(2011)\citenamefont {Helbig},
  \citenamefont {Fuks}, \citenamefont {Tokatly}, \citenamefont {Appel},
  \citenamefont {Gross},\ and\ \citenamefont {Rubio}}]{helbig}%
  \BibitemOpen
  \bibfield  {author} {\bibinfo {author} {\bibfnamefont {N.}~\bibnamefont
  {Helbig}}, \bibinfo {author} {\bibfnamefont {J.}~\bibnamefont {Fuks}},
  \bibinfo {author} {\bibfnamefont {I.}~\bibnamefont {Tokatly}}, \bibinfo
  {author} {\bibfnamefont {H.}~\bibnamefont {Appel}}, \bibinfo {author}
  {\bibfnamefont {E.}~\bibnamefont {Gross}}, \ and\ \bibinfo {author}
  {\bibfnamefont {A.}~\bibnamefont {Rubio}},\ }\href {\doibase
  http://dx.doi.org/10.1016/j.chemphys.2011.06.010} {\bibfield  {journal}
  {\bibinfo  {journal} {Chemical Physics}\ }\textbf {\bibinfo {volume} {391}},\
  \bibinfo {pages} {1 } (\bibinfo {year} {2011})}\BibitemShut {NoStop}%
\bibitem [{\citenamefont {Giesbertz}\ \emph {et~al.}(2012)\citenamefont
  {Giesbertz}, \citenamefont {Gritsenko},\ and\ \citenamefont
  {Baerends}}]{giesbjcp2012}%
  \BibitemOpen
  \bibfield  {author} {\bibinfo {author} {\bibfnamefont {K.~J.~H.}\
  \bibnamefont {Giesbertz}}, \bibinfo {author} {\bibfnamefont {O.~V.}\
  \bibnamefont {Gritsenko}}, \ and\ \bibinfo {author} {\bibfnamefont {E.~J.}\
  \bibnamefont {Baerends}},\ }\href {\doibase
  http://dx.doi.org/10.1063/1.3687344} {\bibfield  {journal} {\bibinfo
  {journal} {The Journal of Chemical Physics}\ }\textbf {\bibinfo {volume}
  {136}},\ \bibinfo {eid} {094104} (\bibinfo {year} {2012})}\BibitemShut
  {NoStop}%
\bibitem{remark} Note that, thanks to the sparsity of $ \gamma_{2,ijkl}$ in \eqref{eq:gamma2tilde}, not all combinations of NO indices contribute.



\end{thebibliography}
\end{document}